\documentclass[reprint,aps,pra]{revtex4-2}

\usepackage{caption}
\usepackage{subcaption}
\usepackage{physics}
\usepackage{amsfonts}
\usepackage{amssymb}
\usepackage{graphicx}
\usepackage{dcolumn}
\usepackage{bm}

\begin{document}

\preprint{Draft - Do Not Distribute}

\title{Entanglement Dynamics of Separable Squeezed States in Finite Memory Structured Reservoir}

\author{Austen Couvertier}
 \altaffiliation[]{acouvert@stevens.edu}
\author{Ting Yu}%
 \email{tyu1@stevens.edu}
\affiliation{Department of Physics, Stevens Institute of Technology
Hoboken, New Jersey 07030, USA
}%

\date{\today}

\begin{abstract}
Entanglement in continuous-variable Gaussian systems is a key resource, and common reservoirs can both suppress and generate correlations. Existing work focused on pre-entangled states or Markovian baths, leaving open whether separable squeezed inputs entangle in structured environments or under modulation. We study two bosonic modes coupled to a common reservoir, each initialized in a separable squeezed vacuum. Dynamics are analyzed utilizing Gaussian covariance methods, evolved under approximate Non-Markovian quantum state diffusion (QSD), finite-temperature pseudomode embeddings, and Bures-based non-Markovian diagnostics. We identify three mechanisms absent in Markovian dynamics: (1) A detuning condition that freezes entanglement trajectories across reservoir correlation times; (2) birth, death, and revival of entanglement from orthogonal inputs; and (3) integer-locked beating with square-wave oscillations produced by periodic detuning. All mechanisms persist at finite temperature, with deviations bounded within $5\%$ in cryogenic regimes and $20\%$ at moderate occupations. These deviation bounds align with cryogenic cavity, phononic, and optomechanical platforms, where structured spectral densities and detuning modulation are already accessible. Structured reservoirs are shown to emerge as tunable entanglement resources for continuous-variable quantum technologies.
\end{abstract}

\maketitle

\section{Introduction}

Entanglement is a central resource in continuous-variable (CV) quantum information, where Gaussian methods provide a complete framework for analysis and applications \cite{adesso_entanglement_2007,weedbrook_gaussian_2012,ferraro_gaussian_2005}. Because CV states couple directly to bosonic environments, dissipation is unavoidable, yet shared baths can also generate correlations. This environment-assisted mechanism was first demonstrated in qubit models \cite{braun_creation_2002,benatti_environment_2003,ficek_entangled_2002} and later extended to Gaussian settings, where two modes coupled to a common reservoir can become entangled \cite{prauzner-bechcicki_two-mode_2004,wolf_entangling_2011,horhammer_environment-induced_2008}. The environment’s dual role as both a noise source and a mediator of correlations remains a defining feature of open-system physics \cite{aolita_open-system_2015,breuer_colloquium_2016,rivas_quantum_2014}. To establish such effects unambiguously, rigorous separability criteria are required, notably the Gaussian form of the Peres–Horodecki test \cite{simon_peres-horodecki_2000} and the Duan-Giedke-Cirac-Zoller variance inequality \cite{duan_inseparability_2000}.

In the Markovian limit, where environmental correlations are memoryless, bosonic modes coupled to a common bath exhibit steady correlations only under aligned squeezing inputs \cite{prauzner-bechcicki_two-mode_2004,horhammer_environment-induced_2008,linowski_stabilizing_2020,croitoru_time_2022}. This outcome is explained by a bright–dark decomposition, in which one collective mode decoheres through the bath while its orthogonal partner remains decoherence-free \cite{lidar_decoherence-free_2003,zhao_dynamics_2015}. Aligned inputs transfer squeezing to the dark mode and yield persistent entanglement, whereas orthogonal inputs decay to separable vacua. Most Markovian analyses instead assume pre-entangled states or introduce direct intermode coupling, leaving separable squeezed states in a shared reservoir unaddressed.

When reservoir correlations persist, non-Markovian dynamics generate behaviors absent in the memoryless limit, including entanglement revivals, oscillatory correlations, and residual steady states \cite{an_non-markovian_2007,vasile_continuous-variable-entanglement_2009,hsiang_entanglement_2022,valido_gaussian_2013,ghesquiere_entanglement_2013}. Most analyses initialize with two-mode squeezed states, leaving open whether separable squeezed inputs can generate correlations under the same conditions. A systematic framework for such dynamics is provided by quantum state diffusion (QSD) and its noise expansion \cite{yu_non-markovian_1999,diosi_non-markovian_1998}, though closures such as the $O_0$ scheme are usually employed as numerical approximations with limited analytic connection between bath parameters and entanglement persistence \cite{barnett_hazards_2001,breuer_colloquium_2016,rivas_quantum_2014,einsiedler_non-markovianity_2020}. Additional evidence comes from Brownian-oscillator models, where low-temperature reservoirs induce residual entanglement through environment-mediated interactions \cite{shiokawa_non-markovian_2009}.

Periodic modulation of couplings has been widely explored as a means to stabilize entanglement. Optomechanical and oscillator models show that amplitude- or frequency-modulated interactions can preserve correlations in noisy environments \cite{bai_amplitude-modulation-based_2019,gonzalez-henao_generation_2015,yang_multi-field-driven_2024}. Reservoir-engineering protocols and multi-tone drives achieve similar control by shaping coupling channels or auxiliary baths \cite{woolley_two-mode_2014,liao_reservoir-engineered_2018}. All of these approaches act on coupling operators, whereas detuning modulation of number operators—analyzed in Sec.~\ref{subsec:beating} and Fig.~\ref{fig:integerLocking}, with derivation in App.~\ref{app:PMA}—remains unexplored in structured reservoirs.

Non-Markovianity is commonly diagnosed through violations of contractivity \cite{breuer_measure_2009,rivas_quantum_2014}. For Gaussian channels, efficient diagnostics are available using the Bures distance, supported by fidelity formulas and error bounds \cite{einsiedler_non-markovianity_2020,banchi_quantum_2015,bittel_optimal_2025}. These witnesses have not been applied to detuning-modulated Gaussian channels, leaving open whether sinusoidal modulation enhances backflow in structured reservoirs or how such effects manifest in Gaussian diagnostics.

We study two bosonic modes in a shared environment, initialized in separable squeezed vacua. Our results reveal three mechanisms not present in the Markovian case: (i) an analytic freezing law that stabilizes correlations against bath-memory variations, (ii) entanglement birth, revival, and long-lived enhancement from orthogonal inputs under detuning, and (iii) periodic beating with integer locking under sinusoidal modulation (Apps.~\ref{app:O0Theory}–\ref{app:PMA}, Figs.~\ref{fig:freezingZeroT}–\ref{fig:integerLocking}). All three persist at finite temperature through pseudomode dynamics (App.~\ref{app:Pseudo}) and are corroborated by Gaussian Bures diagnostics (Fig.~\ref{fig:orthogonalRevivals}(c) and Fig.~\ref{fig:integerLocking}(b)). These effects fall within the experimental reach of atomic-ensemble and optomechanical platforms \cite{krauter_entanglement_2011,ockeloen-korppi_stabilized_2018,park_single-mode_2024,bai_amplitude-modulation-based_2019}, and, to our knowledge, have not been reported previously in CV open systems.

\section{Theoretical Framework}\label{sec:theory}

\subsection{System and Conventions}
\label{sec:sysdef}

We consider two independent bosonic modes with frequencies $\omega_a$ and $\omega_b$, governed by the free Hamiltonian
\begin{equation}\label{eqn:sysHam}
H_S=\hbar\omega_a \hat{a}^\dagger \hat{a} + \hbar\omega_b \hat{b}^\dagger \hat{b}.
\end{equation}
The modes do not interact directly but couple to a common bosonic reservoir through the collective operator $\hat C=\hat a+\hat b$, with the Lindblad jump operator
\begin{equation}\label{eqn:sysJump}
    \hat{L} = \sqrt{\kappa}\,\hat{C}.
\end{equation}
The operator in Eq.~\ref{eqn:sysJump} contains the system decay rate in the real-valued $\kappa$. This form of collective dissipation can generate or suppress entanglement depending on input symmetry \cite{braun_creation_2002,benatti_environment_2003,prauzner-bechcicki_two-mode_2004,horhammer_environment-induced_2008}. We additionally define the mode detuning $\delta_{AB}=\omega_b-\omega_a$ and the system–environment detuning $\delta_{AE}=\Omega-\omega_a$, where $\Omega$ is the central bath frequency.

Initial states are separable Gaussian vacua prepared by single-mode squeezing
\begin{equation}\label{eqn:sysInit}
    \ket{\psi}_0 = \hat{\mathcal{D}}_a(\alpha)\hat{\mathcal{S}}_a(s_a)\ket{0}_a \otimes \hat{\mathcal{D}}_b(\beta)\hat{\mathcal{S}}_b(s_b)\ket{0}_b.
\end{equation}
Displacements are fixed at $\alpha=\beta=\sqrt{5}\,e^{i\pi/4}$ since logarithmic negativity $E_N$ is displacement-invariant. The relevant geometry is the squeezing orientation: aligned inputs place both modes on the same quadrature ($x$ or $p$), while orthogonal inputs place one on $x$ and the other on $p$. Aligned squeezing yields steady correlations in Markovian environments, whereas orthogonal inputs decay to separable vacua unless non-Markovian structure is present \cite{prauzner-bechcicki_two-mode_2004,croitoru_time_2022,horhammer_environment-induced_2008,zhao_two_2003}.

Finally, we adopt a rotating frame at $\omega_a$, measure time in units of $\kappa^{-1}$, frequencies in units of $\kappa$, and therefore set $\kappa=1$ for simplicity. Additionally, we preserve $\hbar$ at the Hamiltonian level while the quadratures are dimensionless by design. This $\hbar$ does not explicitly appear in the rescaled dynamical equations. Under these conventions, the aligned and orthogonal cases establish the baseline against which our non-Markovian results are compared (see App.~\ref{app:BDSS}).

\subsection{Gaussian State Representation}
\label{sec:gaussian}

Bilinear dynamics preserve Gaussianity, so states remain Gaussian under the evolution. Each state is specified by $(\langle \hat R\rangle,\sigma)$, the first moments and covariance matrix, with higher cumulants vanishing. Therefore, the equations of motion for first- and second-order moments completely define the dynamics\cite{adesso_entanglement_2007,weedbrook_gaussian_2012,ferraro_gaussian_2005}. In practice, we evolve a closed system of ladder-operator expectations, equivalent to propagating $(\langle \hat R\rangle,\sigma)$. Quadratures are defined for each mode
\begin{equation}\label{eqn:quadOps}
    \hat{x}_o=\frac{\hat{o}+\hat{o}^\dagger}{\sqrt{2}}, \qquad \hat{p}_o=-\frac{i}{\sqrt{2}}(\hat{o}-\hat{o}^\dagger),
\end{equation}
satisfying $[\hat x_o,\hat p_o]=i$. This recasts operator moments into phase space, the standard entry point for Gaussian analysis. The covariance matrix is defined
\begin{equation}\label{eqn:covMat}
    \sigma_{ij}=\tfrac{1}{2}\langle \{\hat{R}_i-\langle \hat{R}_i\rangle,\,\hat{R}_j-\langle \hat{R}_j\rangle\}\rangle,
\end{equation}
so displacements are admissible. The first moments vanish under logarithmic negativity but are retained under fidelity and Bures diagnostics. In this work, the mean vector and covariance matrix are defined via quadrature ordering
\begin{equation}\label{eqn:quadOrder}
   \hat{R} = (\hat{x}_{a},\hat{x}_{b},\hat{p}_{a},\hat{p}_{b}).
\end{equation}
Utilizing the mode-level quadrature commutators, the symplectic form is defined
\begin{equation}\label{eqn:sympForm}
\mathbf{J}=\begin{pmatrix}0&1\\-1&0\end{pmatrix} \otimes \begin{pmatrix}1&0\\0&1\end{pmatrix},
\end{equation}
and the bona fide condition for a physical Gaussian state is $\sigma+i\mathbf{J}/2\succeq0$, equivalent to the Simon and Duan separability tests \cite{simon_peres-horodecki_2000,duan_inseparability_2000}. These conventions underpin all subsequent use of symplectic eigenvalues and logarithmic negativity (App.~\ref{app:BDSS}) and provide reproducible ordering checks for numerical analysis (App.~\ref{app:Numerics}).

\subsection{Entanglement and Fidelity Measures}
\label{sec:measures}

Entanglement is measured with logarithmic negativity, a computable Gaussian monotone that captures transient and steady correlations \cite{adesso_entanglement_2007,weedbrook_gaussian_2012,simon_peres-horodecki_2000,duan_inseparability_2000,vidal_computable_2002}. For the quadrature ordering of the bipartite covariance matrix, partial transposition is done by flipping the sign of mode $A$’s momentum quadrature with $P_a=\mathrm{diag}(1,1,-1,1)$. The smallest symplectic eigenvalue $\tilde\nu_-=\min(\mathrm{eig}(|i\mathbf{J} P_a\sigma P_a|))$ sets the separability. The measure is defined
\begin{equation}\label{eq:EN}
    E_N=\max\left[0,-\ln\left(2\tilde{\nu}_-\right)\right].
\end{equation}
Fidelity gives the overlap between two Gaussian states. We use the root-Uhlmann fidelity, which admits a closed-form expression in phase space\cite{banchi_quantum_2015}. It depends on covariances and displacements as
\begin{equation}\label{eqn:Fid}
    \mathrm{F}(\rho_1,\rho_2) = \mathrm{F}_{0}(\sigma_1,\sigma_2)e^{-\frac{1}{4}(\langle \hat R_{2}\rangle - \langle \hat R_{1}\rangle)^{T}(\sigma_1+\sigma_2)^{-1}(\langle \hat R_{2}\rangle - \langle \hat R_{1}\rangle)}.
\end{equation}
Here $F_0(\sigma_1,\sigma_2)$ is written with determinants of $\sigma_1$, $\sigma_2$, $\sigma_1+\sigma_2$, and $\mathbf{J}$. Fidelity is defined for two states $\rho_i=(\langle \hat R_i\rangle,\sigma_i)$ with $i\in \{1,2\}$. State separation is derived from this Uhlmann root-fidelity to define the Bures distance as
\begin{equation}\label{eqn:Bures}
D_B(\rho_1,\rho_2)=\sqrt{2\left(1-\mathrm{F}(\rho_1,\rho_2)\right)}. 
\end{equation}
This makes it a standard witness of non-Markovianity in Gaussian channels.

In what follows, logarithmic negativity tracks bipartite correlations, while the Bures framework tests backflow. Both are taken from the covariance matrix defined in Sec.~\ref{sec:gaussian}. Their consistent use is checked in App.~\ref{app:Numerics}.

\subsection{Non-Markovianity Witness}
\label{sec:nonmarkov}

In a CP-divisible channel, any contractive metric decreases monotonically. The Bures distance between two evolving states is such a probe. A temporary increase, $\dot D_B(t)>0$, signals backflow and memory effects in the environment \cite{breuer_measure_2009,rivas_quantum_2014}.

We quantify this behavior with the integrated witness
\begin{equation}\label{eqn:N}
    \mathcal N=\int_{\dot D_B>0}\dot D_B(t)\,dt \;\;\approx \sum_{\Delta D_B>0}\Delta D_B.
\end{equation}
The derivative $\dot D_B(t)$ is approximated by a discrete sum over positive increments, $\Delta D_B=D_B(t_{i+1})-D_B(t_i)$. Our implementation uses antipodal Gaussian states. They share the same covariance matrix but have opposite mean displacements. This choice is consistent with Sec.~\ref{sec:gaussian}, treats displacements explicitly, and increases sensitivity without requiring full input optimization \cite{einsiedler_non-markovianity_2020}. Numerical evaluation follows the procedure in App.~\ref{app:Numerics}.

This measure is only a witness. $\mathcal N>0$ confirms non-Markovianity. $\mathcal N=0$ does not exclude memory, only that this probe does not detect it. Large-detuning regimes can still support finite entanglement while the witness vanishes, as shown in Figs.~\ref{fig:orthogonalRevivals}c and \ref{fig:integerLocking}b.

\subsection{Perturbative Non-Markovian QSD Formalism}
\label{sec:oZero}

In general, we consider the system defined in Eq.~\ref{eqn:sysHam} to be coupled linearly to an infinite collection of bosonic modes. The coupling is provided by the operator in Eq.~\ref{eqn:sysJump} and defines the environmental dynamics as
\begin{equation}\label{eqn:envHam}
    \hat{H}_{E} = \hbar\sum_{\bm{\lambda}}\omega_{\bm{\lambda}}\hat{a}^{\dagger}_{\bm{\lambda}}\hat{a}_{\bm{\lambda}} + g_{\bm{\lambda}}(\hat{a}_{\bm{\lambda}}\hat{L}^{\dagger}+\hat{a}^{\dagger}_{\bm{\lambda}}\hat{L}).
\end{equation}
In this Hamiltonian, $\hat{a}_{\bm{\lambda}}$, $\omega_{\bm{\lambda}}$, and $g_{\bm{\lambda}}$ correspond to the ladder operator, mode-frequency, and system-environment coupling strength of the environment. Using Eqs.~\ref{eqn:sysHam} and \ref{eqn:envHam} the total dynamics are described by $\hat{H}_{\mathrm{tot}} = \hat{H}_{E} + \hat{H}_{S}$. At present, we assume the collection of bosonic modes is at zero temperature. In the linear interaction case, the influence of the environment is encoded in the distribution of the couplings, $g_{\bm{\lambda}}$, which a bath correlation function can specify. Our work models this correlation by the Ornstein–Uhlenbeck kernel
\begin{equation}\label{eqn:kernel}
    \alpha(t-s)=\tfrac{\gamma}{2}e^{-(\gamma+i\Omega)(t-s)},
\end{equation}
characterized by a decay rate $\gamma$ and central frequency $\Omega$. Its finite correlation time introduces memory absent in the Markovian limit.

To incorporate such correlations in QSD, we use the $O_0$ operator closure \cite{yu_non-markovian_1999,diosi_non-markovian_1998,barnett_hazards_2001}. The ansatz restricts $\hat O_0(t,s)$ to a linear combination of the annihilation operators
\begin{equation}\label{eqn:oZeroAnsatz}
    \hat{O}_{0}(t,s)=f_{1}(t,s)\hat{a}+f_{2}(t,s)\hat{b}.
\end{equation} 
Averaging gives a time-local form
\begin{equation}\label{eqn:oBar}
    \hat{\bar{O}}_{0}(t)=F_{1}(t)\hat{a}+F_{2}(t)\hat{b},
\end{equation}
with convolution coefficients $F_i(t)=\int_0^t\alpha(t,s)f_i(t,s)ds$. These encode memory and evolve by consistency relations. The construction preserves Gaussianity and yields closed dynamics for system observables.

The closure enters the Heisenberg equations through a modified master equation
\begin{equation}\label{eqn:oZeroEOM}
    \frac{d}{dt}\langle \hat{A} \rangle = -\frac{i}{\hbar}\langle[\hat{A},\hat{H}_{S}]\rangle + \langle \hat{\bar{O}}_{0}^{\dagger}(t)[\hat{A},\hat{L}] + [\hat{L}^{\dagger},\hat{A}]\hat{\bar{O}}_{0}(t) \rangle,
\end{equation}
combining Hamiltonian terms with feedback via $\hat{\bar O}_0(t)$. Projecting onto the annihilation operators leads to coupled Riccati equations for $F_1$ and $F_2$
\begin{equation}\label{eqn:consistency}
    \begin{aligned} 
\frac{d}{dt}F_{1}&=  \frac{\gamma\sqrt{\kappa}}{2} + (\sqrt{\kappa}(F_1 + F_2)-(\gamma+i\delta_{AE}))F_1\\ 
\frac{d}{dt}F_{2}&= \frac{\gamma\sqrt{\kappa}}{2} + (\sqrt{\kappa}(F_1 + F_2)-(\gamma+i(\delta_{AE}-\delta_{AB})))F_2. \nonumber
    \end{aligned}
\end{equation}
They include a source term $\gamma\sqrt{\kappa}/2$ for reservoir exchange, nonlinear damping $\sqrt{\kappa}(F_1+F_2)$, and detuning shifts $\delta_{AE},\delta_{AB}$. Initial conditions $F_1(0)=F_2(0)=0$ ensure correlations build only after coupling begins. Numerical integration propagates $F_{1,2}(t)$, which determines quadrature covariances and feeds into entanglement and fidelity measures.

For resonant modes ($\delta_{AB}=0$), the coefficients converge to equal steady values whenever $\delta_{AE} \neq 0$ or $\gamma /\kappa \geq 4$ as
\begin{equation}\label{eqn:modeResonanceSS}
    F_{1,2}(\infty)=\lim_{t\to\infty}F_{1,2}(t) = \frac{1}{4\sqrt{\kappa}}(\gamma + i\delta_{AE} - \chi(\gamma,\delta_{AE})),
\end{equation}
with parameter:
\begin{align}\label{eqn:SSFuncts}
    \chi(\gamma,\delta_{AE})&=\sqrt{(\gamma+i\delta_{AE})^{2}-4\gamma\kappa}.
\end{align}
The real part of $F_{1,2}(\infty)$ sets the damping, and the imaginary part sets the oscillation frequency. The principal branch of $\chi$ is chosen with $\Re \chi>0$ to ensure stability. We also emphasize that $F_{1,2}(\infty)$ is distinct from the usual Markovian limit, $\gamma \to \infty$, which would cause $F_{1,2}(\infty)$ to further simplify. Specifically, $\lim_{\gamma \to \infty}F_{1,2}(t) =\sqrt{\kappa}/2$, which recovers the Markovian dynamics exactly and is distinct from Eq.~\ref{eqn:modeResonanceSS} unless $\gamma/\kappa \gg 4$.

Two decay scales characterize relaxation: $(t_s,n_s)$ for coefficient decay and $(t_n,n)$ for second-order moments. These definitions, detailed in App.~\ref{app:Numerics}, enable comparison of memory across reservoirs. From the analytic $F_{1,2}(\infty)$, one derives inequalities: the critical detuning $\delta_{AE}^*$,
\begin{equation}\label{eqn:environmentFreezing}
     \delta_{AE}^{*}=\frac{\sqrt{\,2\kappa t_{n}\!\left[(2t_{n}\gamma-\ln n)^{-1}+(\ln n)^{-1}\right]-1\,}\,(t_{n}\gamma-\ln n)}{t_{n}},
\end{equation}
and a memory bound on $\gamma$,
\begin{equation}\label{eqn:freezingMemoryBound}
     \gamma \;\geq\; \frac{\ln n}{t_{n}}+\frac{\ln n_{s}}{t_{s}}.
\end{equation}
Under these conditions, trajectories collapse onto a common curve independent of $\gamma$, a freezing behavior linked to non-Markovian stabilization \cite{maniscalco_entanglement_2007,vasile_continuous-variable-entanglement_2009}.

These derivations are given in App.~\ref{app:O0Theory}. The inequalities define the freezing law used to interpret entanglement dynamics in Sec.~\ref{subsec:freezing} \cite{breuer_measure_2009,rivas_quantum_2014,einsiedler_non-markovianity_2020,hesabi_memory_2023}.

\subsection{Finite Temperature Pseudomode Formalism}
\label{sec:pseudo}

The $O_0$ closure of Sec.~\ref{sec:oZero} applies only at zero temperature. At finite $T$, thermal absorption processes can violate complete positivity when treated within the same ansatz. To preserve physicality, we embed the Ornstein–Uhlenbeck reservoir Eq.~\ref{eqn:kernel} into a Markovian setting by introducing a single auxiliary bosonic mode $\hat c$ of frequency $\Omega$. It couples linearly to the collective operator of Eq.~\ref{eqn:sysJump}, giving the effective Hamiltonian
\begin{equation} \label{eqn:pseudoHam}
\hat{H}_{\mathrm{eff}} = \hat{H}_{S} + \hbar\Omega \hat{c}^\dagger \hat{c} + \hbar g(\hat{L}\hat{c}^\dagger + \hat{L}^\dagger \hat{c}),
\end{equation}
with $g=\sqrt{\gamma/2}$ chosen to reproduce the target correlation. This maps the non-Markovian bath to a Markovian tripartite system, enabling consistent treatment of finite-$T$ noise \cite{pleasance_generalized_2020,horhammer_environment-induced_2008,shiokawa_non-markovian_2009}.

Finite temperature is included by coupling the pseudomode to a memoryless background through emission and absorption channels
\begin{equation}\label{eqn:pseudoJumps}
    \hat{L}_{-} = \sqrt{2\gamma(\bar{n}+1)}\,\hat{c}, \qquad \hat{L}_{+} = \sqrt{2\gamma \bar{n}}\,\hat{c}^{\dagger},
\end{equation}
where $\bar n=[e^{\hbar\Omega/k_BT}-1]^{-1}$ is the mean thermal photon occupation. With this choice, $\langle \hat c\rangle$ decays at rate $\gamma$, so correlations match the Ornstein–Uhlenbeck form. The pseudomode is initialized in its thermal steady state with occupation $\bar n$ to avoid transients.

The enlarged system $(a,b,c)$ evolves under the Gorini--Kossakowski--Lindblad--Sudarshan (GKLS) master equation\cite{breuer_theory_2009, lindblad_generators_1976}
\begin{equation}\label{eqn:pseudoEOM}
     \frac{d}{dt}\langle \hat{A} \rangle = -\frac{i}{\hbar}\langle[\hat{A},\hat{H}_{\mathrm{eff}}]\rangle + \frac{1}{2}\sum_{i\in[-,+]}\langle \hat{L}^{\dagger}_{i}[\hat{A},\hat{L}_{i}] +  [\hat{L}^{\dagger}_{i},\hat{A}]\hat{L}_{i} \rangle.
\end{equation}
Because the embedding is Markovian, the evolution is completely positive for all $t$, and tracing out $c$ yields CP reduced dynamics for $(a,b)$. This contrasts with the zero-$T$ closure, where positivity is not automatic.

After evolution, $c$ is traced out to obtain reduced covariances for $a$ and $b$. In the limit $\bar n=0$, the pseudomode embedding reproduces the zero-$T$ Ornstein–Uhlenbeck kernel and $\hat{\bar O}_0$ dynamics of Sec.~\ref{sec:oZero}. For $\bar n>0$, it gives a controlled extension that incorporates both emission and absorption.

This construction links zero-temperature QSD closure with a Markovian embedding valid at all $T$, equivalent to other pseudomode approaches in non-Markovian theory \cite{barnett_hazards_2001,pleasance_generalized_2020,luo_quantum-classical_2023,garraway_nonperturbative_1997,maniscalco_entanglement_2007,vasile_continuous-variable-entanglement_2009}. Full derivations are in App.~\ref{app:Pseudo}. It allows finite-temperature dynamics to be analyzed on the same footing as the zero-temperature case, forming the basis for Secs.~\ref{subsec:freezing},~\ref{subsec:beating}.
 
\subsection{Numerical Protocol}\label{sec:numerical}

We chose an adaptive explicit Runge–Kutta method with automatic stiff fallback, using \texttt{MachinePrecision} as the threshold. ODEs were solved adaptively and results sampled on a uniform grid with $\kappa\Delta t=0.01$ up to $\kappa t=150$. Quantities that depend on derivatives use the solver interpolation. Error control followed Wolfram’s rule: $\mathrm{error} <10^{-\texttt{AccuracyGoal}}+|x|*10^{-\texttt{PrecisionGoal}}$. We set \texttt{AccuracyGoal = PrecisionGoal = 14}. Convergence was tested by halving the step size and tightening the goals by +2, requiring $\max_t|\Delta E_N|<10^{-6}$, $\max_t|\Delta D_B|<10^{-6}$, and an integrated change $<10^{-5}$. These conditions were satisfied at \texttt{MachinePrecision}.

Simulations were in the rotating frame at $\omega_a$ with $\kappa=1$, giving dimensionless quantities. Production runs used $|s|\leq1$; exploratory scans extended to $|s|\leq5$. Bath parameters covered $\gamma/\kappa\in[0.5,5]$, with a Markov check at $\gamma/\kappa=10$ (App.~G). Detunings were sampled as $\delta_{AB}/\kappa\in[-2,0]$ and $\delta_{AE}/\kappa\in\{0.1,1,10\}$, using symmetry for $|\delta_{AE}/\kappa|>1$. Thermal occupations were taken over $0\leq\bar n\leq1$.

At each step we enforced Gaussian physicality: $\sigma+i\mathbf{J}/2\succeq0$, symmetry of $\sigma$ to $10^{-10}$, minimum symplectic eigenvalue $\tilde\nu_{\min}\geq1/2-10^{-4}$, and all eigenvalues $\geq10^{-4}$ \cite{adesso_entanglement_2007,weedbrook_gaussian_2012}. Any violation marked the parameter point unstable.

For the Bures witness, we built antipodal Gaussian pairs with identical $\sigma(t)$ but opposite mean vectors. Both were then evolved under identical system conditions, and fidelity and Bures distance were computed as defined in Secs.~\ref {sec:measures}. More details on physicality and convergence are in App.~\ref{app:Numerics}. Zero-temperature correspondence is given in App.~\ref{app:Pseudo}.

These protocols ensure that all reported trajectories are physically valid and converge within the stated tolerances.

\section{Results}\label{sec:results}

\subsection{Markov Baseline}\label{subsec:markovBaseline}

\subsubsection{Zero-Temperature entanglement trajectories}

\begin{figure}
    \centering
    \includegraphics[width=\linewidth]{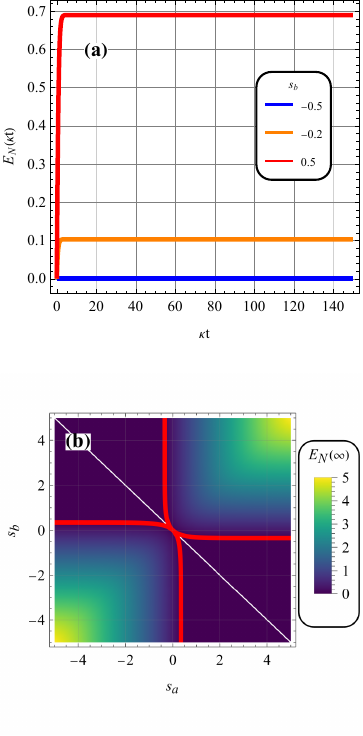}
    \caption{(a) Time evolution of the logarithmic negativity $E_N$ [Eq.~\ref{eq:EN}] in a Markovian bath with mode $A$ fixed at $s_a=+1$ and mode $B$ initialized at $s_b=-0.5$ (blue), $s_b=-0.2$ (orange), and $s_b=+0.5$ (red). (b) Steady-state $E_N$ as a function of $(s_a,s_b)$ over $[-5,5]$, with bright regions bounded by Simon criteria [Eqs.~\ref{eqn:SimonPos}-\ref{eqn:SimonNeg}].}
    \label{fig:markovBaseline}
\end{figure}

We first examine the Markovian limit in which two uncoupled bosonic modes couple symmetrically to a memoryless reservoir, with entanglement quantified by the logarithmic negativity defined in Eq.~\ref{eq:EN}. Figure~\ref{fig:markovBaseline}(a) shows trajectories of $E_N(t)$ for mode $A$ fixed at $s_a=+1$ and mode $B$ prepared with $s_b=-0.5$ (blue), $s_b=-0.2$ (orange), and $s_b=+0.5$ (red). Aligned inputs lead to rapid approach to a finite steady entanglement, weakly orthogonal inputs yield only a small plateau, and strongly orthogonal inputs remain separable. The heat map in Fig.~\ref{fig:markovBaseline}(b) confirms that steady entanglement is restricted to the aligned quadrants, with bright regions bounded by the Simon criteria of Eqns.~\ref{eqn:SimonPos}-\ref{eqn:SimonNeg}~\cite{simon_peres-horodecki_2000,vidal_computable_2002,adesso_entanglement_2007,ferraro_gaussian_2005}.

The mechanism can be understood in the bright--dark mode basis, where the collective channel acts only on the bright mode and leaves the dark mode decoherence-free~\cite{lidar_decoherence-free_2003,zhao_density_2002,chou_exact_2008}. When both inputs are aligned, squeezing is distributed across bright and dark modes so that the long-time state consists of a vacuum in the bright mode and a squeezed dark mode, which transforms back into an entangled steady state in the original basis~\cite{adesso_entanglement_2007,ferraro_gaussian_2005}. For orthogonal inputs, only the bright mode is initially squeezed, yielding separable vacua. This structure is captured analytically by the symplectic eigenvalues and Simon bounds derived in Appendix~\ref{app:BDSS}. The example presented in Appendix~\ref{app:BDSS} shows for $s_a=1$, the bound requires $s_b>-0.312$, matching the numerical threshold in Fig.~\ref{fig:markovBaseline}(a).

\subsubsection{Thermal suppression of steady state entanglement}

\begin{figure}
    \centering
    \includegraphics[width=\linewidth]{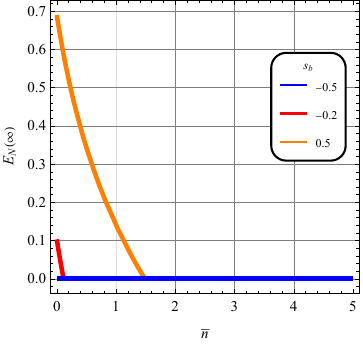}
    \caption{Steady-state logarithmic negativity $E_N$ [Eq.~\ref{eq:EN}] versus thermal occupation $\bar n$ in a Markovian bath. Results are shown for aligned ($s_a=s_b=+0.5$, orange), weakly orthogonal ($s_a=+0.5$, $s_b=-0.2$, red), and strongly orthogonal ($s_a=+0.5$, $s_b=-0.5$, blue) inputs. Aligned inputs retain finite entanglement up to $\bar n\!\approx\!1.5$, weakly orthogonal inputs separate for $\bar n \approx 0.2$, and strongly orthogonal inputs remain separable for all $\bar n$.}
    \label{fig:markovThermal}
\end{figure}

Including finite thermal occupation rescales the partial-transpose symplectic spectrum by $\sqrt{1+2\bar n}$, thereby tightening the separability condition and suppressing steady correlations. Figure~\ref{fig:markovThermal} shows that aligned inputs retain nonzero entanglement up to $\bar n\!\approx\!1.5$, while weakly orthogonal inputs become separable at $\bar n\!\approx\!0.2$. Strongly orthogonal inputs remain separable for all $\bar n$. This monotonic degradation is consistent with prior studies of thermal suppression in Gaussian channels~\cite{dumitru_entanglement_2015,ghesquiere_entanglement_2013,horhammer_environment-induced_2008}. Our extension relies on the pseudomode embedding formalism~\cite{pleasance_generalized_2020,luo_quantum-classical_2023}, ensuring complete positivity at $\bar n>0$. Thus, the Markov baseline demonstrates that steady entanglement survives only for aligned squeezing and only within cryogenic thermal windows.

\subsection{Freezing under Structured Environments}\label{subsec:freezing}

\subsubsection{Zero-temperature freezing law}

\begin{figure}
    \centering
    \includegraphics[width=\linewidth]{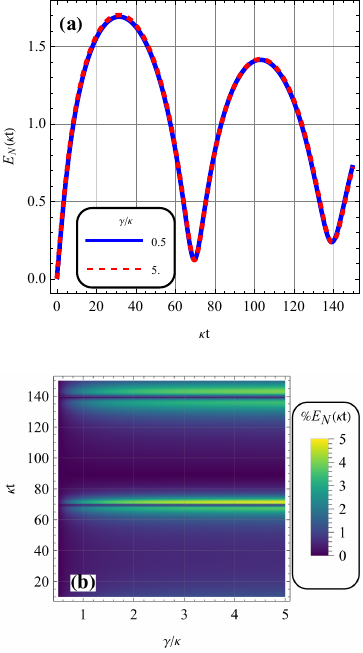}
    \caption{(a) Time evolution of the logarithmic negativity $E_N$ [Eq.~\ref{eq:EN}] in an Ornstein--Uhlenbeck bath for $\gamma/\kappa=0.5$ (blue) and $\gamma/\kappa=5$ (red) at fixed detuning $\delta_{AE}^*$. Trajectories overlap within a few percent. (b) Relative deviation $\%E_N=|\Delta E_N/E_N^{\mathrm{baseline}}|\times100\%$ versus $\gamma$, referenced to the $\gamma/\kappa=0.5$ curve. Deviations remain below $5\%$, concentrated near oscillation minima.}
    \label{fig:freezingZeroT}
\end{figure}

We next consider an Ornstein--Uhlenbeck reservoir at resonance, where memory effects are controlled by the correlation time $1/\gamma$. Figure~\ref{fig:freezingZeroT}(a) compares the entanglement trajectories for $\gamma/\kappa=0.5$ (blue) and $\gamma/\kappa=5$ (red). Despite a tenfold change in memory scale, the curves nearly overlap, oscillating between $E_N\!\approx\!0.2$ and $E_N\!\approx\!1.7$ around the Markovian steady value of unity. The key observation is that the entanglement dynamics remain effectively insensitive to the bath memory, a feature we identify as \emph{freezing} of the structured environment's response.

This insensitivity can be traced to the convergence of the convolution coefficients $F_{i}(t)$ defined in Eq.~\ref{eqn:oBar}, which approach the universal steady value given by Eq.~\ref{eqn:modeResonanceSS}. The analytic freezing criteria of Eqs.~\ref{eqn:environmentFreezing}--\ref{eqn:freezingMemoryBound} guarantee that once these coefficients saturate, the subsequent dynamics collapse onto the Markovian trajectory regardless of $\gamma$. Figure~\ref{fig:freezingZeroT}(b) confirms this prediction: relative deviations remain below $5\%$, concentrated at oscillation minima. These findings align with earlier analyses of QSD closures~\cite{yu_non-markovian_1999,barnett_hazards_2001} and memory-assisted stabilization~\cite{hesabi_memory_2023}.

\subsubsection{Finite-temperature robustness}

\begin{figure}
    \centering
    \includegraphics[width=\linewidth]{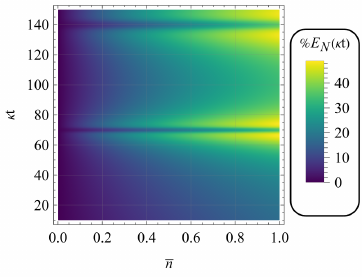}
    \caption{Relative deviation $\%E_N=|\Delta E_N/E_N^{\mathrm{baseline}}|\times100\%$ of freezing dynamics versus thermal occupation $\bar n$. Deviations are bounded by $\leq 5\%$ in the cryogenic regime ($\bar n\leq 0.05$) and $\leq 20\%$ at moderate occupancies ($\bar n\leq 0.2$). Larger excursions occur only at short-time transients and late-time decay.}
    \label{fig:freezingThermal}
\end{figure}

To assess stability against thermal noise, we embed the Ornstein--Uhlenbeck kernel into a pseudomode representation that extends the $\hat O_{0}$ closure to finite temperature~\cite{pleasance_generalized_2020,luo_quantum-classical_2023,shiokawa_non-markovian_2009}. Figure~\ref{fig:freezingThermal} shows the relative deviation of $E_N(\kappa t)$ as a function of bath occupation $\bar n$. Deviations remain within $5\%$ for cryogenic $\bar n\leq0.05$ and within $20\%$ for moderate $\bar n\leq0.2$, while larger excursions occur only at early-time transients and late-time decay. Utilizing the Bose-Einstein definition in Sec.~\ref{sec:pseudo}, a representative mode frequency of $\Omega/2\pi=10~\mathrm{MHz}$ corresponds to temperatures of approximately $158~\mu\mathrm{K}$ and $268~\mu\mathrm{K}$, respectively. For a representative microwave frequency $\Omega/2\pi=10~\mathrm{GHz}$, the corresponding temperatures are $158~\mathrm{mK}$ and $268~\mathrm{mK}$.

Importantly, the oscillatory freezing structure persists across the full range of occupations considered, including $\bar n=1$, corresponding to approximately $693~\mu\mathrm{K}$ at $10~\mathrm{MHz}$ and $693~\mathrm{mK}$ at $10~\mathrm{GHz}$. For microwave CV realizations, this temperature scale lies well within the range accessible to dilution-refrigerator experiments.~\cite{enss_low-temperature_2005,krinner_engineering_2019} For MHz mechanical or motional realizations, however, sub-millikelvin effective occupations generally require active cooling rather than passive thermalization of the bath~\cite{lucas_quantum_2023,serafini_gaussian_2020}. Thus, the relevant experimental interpretation is platform-dependent: the finite-temperature robustness is directly compatible with microwave cryogenic operation, whereas on MHz platforms it corresponds to effectively cooled occupations achievable via reservoir engineering, sideband cooling, or related preparation techniques. Taken together, these results indicate that the environmental stabilization mechanism persists under thermally relevant and experimentally accessible conditions.

\subsection{Orthogonal Squeezing: Revival and Enhancement}\label{subsubsection:modeDetuning}

\begin{figure*}
    \centering
    \includegraphics[width=\linewidth]{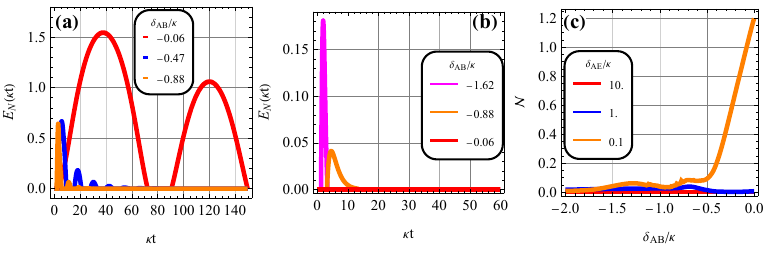}
    \caption{(a) Time evolution of logarithmic negativity $E_N$ [Eq.~\ref{eq:EN}] for orthogonal inputs ($s_a=+1$, $s_b=-1$) in structured baths with different detuning pairs $(\delta_{AB},\delta_{AE})/\kappa$, showing entanglement birth, death, and revival. (b) Markovian comparison, where no choice of $\delta_{AE}$ yields revivals and correlations decay rapidly. (c) Non-Markovianity measure $\mathcal{N}$ [Eq.~\ref{eqn:N}] versus $\delta_{AB}/\kappa$ for the same conditions; in the large-$\delta_{AE}/\kappa$ regime the witness remains zero despite nontrivial $E_N(t)$.}
    \label{fig:orthogonalRevivals}
\end{figure*}

\subsubsection{Revival and long-lived correlations in structured baths}

We now initialize the system with orthogonal squeezing, $s_a=+1$ and $s_b=-1$, and examine the dynamics in an Ornstein--Uhlenbeck reservoir. Figure~\ref{fig:orthogonalRevivals}(a) shows that the entanglement response depends sensitively on detuning. For $(\delta_{AB},\delta_{AE})/\kappa=(-0.88,0.1)$ the logarithmic negativity peaks near $0.65$ before vanishing by $\kappa t\!\approx\!20$, while $(\delta_{AB},\delta_{AE})/\kappa=(-0.47,1)$ produces several small revivals before complete decay by $\kappa t\!\approx\!60$. Near resonance, $(\delta_{AB},\delta_{AE})/\kappa=(-0.06,10)$ yields a maximum $E_N\!\approx\!1.5$ and long-lived revivals around $1.1$. These behaviors arise from detuning-induced phase accumulation $e^{\pm i\delta_{AB}t}$ combined with OU-kernel feedback in the coefficient dynamics [Eq.~\ref{eqn:consistency}, Appendix~\ref{app:O0Theory}], which preserves dark-mode squeezing and enables sustained correlations~\cite{prauzner-bechcicki_two-mode_2004,horhammer_environment-induced_2008,linowski_stabilizing_2020}.

\subsubsection{Markovian baseline comparison}

For comparison, Fig.~\ref{fig:orthogonalRevivals}(b) shows the same orthogonal input evolved in a memoryless reservoir. In this case, the entanglement dynamics reduce to brief birth--death events. At $\delta_{AB}/\kappa=-1.62$ a small peak of $E_N\!\approx\!0.17$ appears near $\kappa t\!\approx\!2$ before vanishing within four time units, while at $\delta_{AB}/\kappa=-0.88$ the response is limited to a minor plateau of $0.04$ between $\kappa t=4$--20. Near resonance, $\delta_{AB}/\kappa=-0.06$, no correlations develop at all. The Markovian master equation with bright--dark decomposition [Eq.~\ref{eqn:BDHam}]  predicts exactly this short-lived behavior, since detuning mixes the modes but without OU-kernel feedback all correlations rapidly decay~\cite{prauzner-bechcicki_two-mode_2004,horhammer_environment-induced_2008,braun_creation_2002,benatti_environment_2003}.

\subsubsection{Non-Markovian witness and entanglement mismatch}

The integrated witness $\mathcal N$ defined in Eq.~\ref{eqn:N} provides a complementary probe of environmental memory. Figure~\ref{fig:orthogonalRevivals}(c) shows its dependence on mode detuning. For $\delta_{AE}/\kappa=10$ the witness is identically zero, even though Fig.~\ref{fig:orthogonalRevivals}(a) exhibits substantial and long-lived entanglement. At $\delta_{AE}/\kappa=1$ the witness becomes finite only for $\delta_{AB}/\kappa<-0.4$, peaking near $0.1$, while for $\delta_{AE}/\kappa=0.1$ a maximum appears close to resonance. These results highlight that $\mathcal N$ detects certain channels of information backflow but does not coincide with the revival structure of $E_N(\kappa t)$, underscoring the limitations of backflow witnesses in capturing environment-assisted entanglement~\cite{breuer_measure_2009,rivas_quantum_2014,einsiedler_non-markovianity_2020,aolita_open-system_2015}.

\subsection{Periodic Modulation: Beating and Integer Locking}\label{subsec:beating}

\subsubsection{Beating dynamics at integer modulation amplitudes}

\begin{figure}
    \centering
    \includegraphics[width=\linewidth]{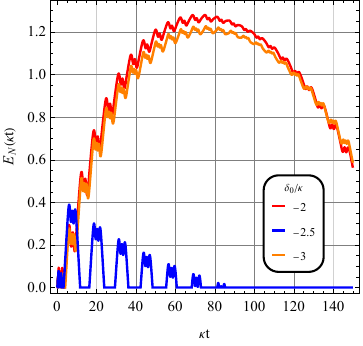}
    \caption{Logarithmic negativity $E_N$ [Eq.~\ref{eq:EN}] under sinusoidal detuning modulation $\delta_{AB}(t)=\delta_0(\sin t+1)$. Integer values of $\delta_0/\kappa$ yield long-lived beating envelopes with square-wave oscillations, while non-integer values suppress sustained correlations.}
    \label{fig:integerBeating}
\end{figure}

We now introduce a sinusoidal modulation of the mode detuning, $\delta_{AB}(t)=\delta_0(\mathrm{sin}(t)+1)$, applied to the orthogonal input state. Figure~\ref{fig:integerBeating} demonstrates that robust beating envelopes appear only when the modulation amplitude takes integer values. For $\delta_0/\kappa=-2$, the logarithmic negativity exhibits persistent oscillations with two square-wave spikes per cycle, maintaining values above $0.6$ even at late times. At $\delta_0/\kappa=-3$, the envelope displays three spikes per cycle with maxima exceeding unity. In contrast, non-integer driving, exemplified by $\delta_0/\kappa=-2.5$, produces destructive interference and rapid decay of correlations. These behaviors are explained by the Jacobi--Anger expansion of the accumulated phase [Appendix~\ref{app:PMA}], which yields constructive interference only when $\delta_0/\omega\in \mathbb{Z}$. Integer modulation uniquely stabilizes long-lived entanglement through harmonic selection of revival pathways~\cite{aolita_open-system_2015,hsiang_entanglement_2022,yang_quantum_2025}.

\subsubsection{Integer locking of time-averaged entanglement}

\begin{figure}
    \centering
    \includegraphics[width=\linewidth]{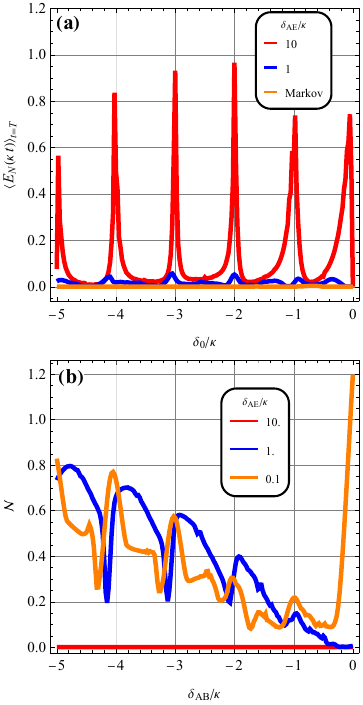}
    \caption{(a) Time-averaged logarithmic negativity $\langle E_N\rangle_T=\tfrac{1}{T}\int_0^T E_N(t)\,dt$ [Eq.~\ref{eq:EN}] versus modulation amplitude $\delta_0/\kappa$ for structured baths with different $\delta_{AE}/\kappa$, showing sharp peaks at integer values. The Markovian baseline remains near zero. (b) Non-Markovianity measure $\mathcal{N}$ [Eq.~\ref{eqn:N}] versus $\delta_{0}/\kappa$ under the same conditions, exhibiting quasi-periodic oscillations; entanglement peaks coincide with transitions rather than maxima in $\mathcal{N}$.}
    \label{fig:integerLocking}
\end{figure}

The long-time average of the logarithmic negativity, defined as $\langle E_N\rangle_T = \tfrac{1}{T}\int_0^T E_N(t)\,dt$, provides a global measure of modulation-induced stabilization. Figure~\ref{fig:integerLocking}(a) shows that for large environment detuning ($\delta_{AE}/\kappa=10$) this average develops sharp peaks at integer $\delta_0/\kappa$ values between $-1$ and $-4$, consistent with the Kronecker-delta selection predicted by the Jacobi--Anger expansion [Appendix~\ref{app:PMA}]. At smaller detuning ($\delta_{AE}/\kappa=1$), the peaks shift off-integer and are suppressed to $\langle E_N\rangle_T\leq 0.05$, while in the Markovian baseline, the response remains near zero. These results confirm that periodic driving locks entanglement to discrete modulation amplitudes, a mechanism absent in memoryless reservoirs~\cite{yu_non-markovian_1999,barnett_hazards_2001,linowski_stabilizing_2020}.

\subsubsection{Backaction response under modulation}

The modulation-induced memory is further assessed using the integrated witness $\mathcal N$ defined in Eq.~\ref{eqn:N}. Figure~\ref{fig:integerLocking}(b) shows that for $\delta_{AE}/\kappa=10$ the witness vanishes identically, even in regimes where Fig.~\ref{fig:integerLocking}(a) reveals strong integer-locked entanglement. At moderate detuning, $\delta_{AE}/\kappa=1$, $\mathcal N$ exhibits quasi-periodic oscillations, reaching maxima below integer values of $\delta_0/\kappa$ and minima above them, while for $\delta_{AE}/\kappa=0.1$ it revives near resonance. Thus, entanglement maxima coincide with transition points of $\mathcal N$ rather than its peaks. This mismatch illustrates that backflow diagnostics based on Bures distance~\cite{breuer_measure_2009,rivas_quantum_2014,einsiedler_non-markovianity_2020} do not capture the stabilization mechanism responsible for periodic beating.

\subsubsection{Finite-temperature persistence of beating}

\begin{figure}
    \centering
    \includegraphics[width=\linewidth]{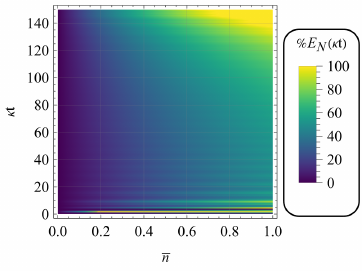}
    \caption{Relative deviation $\%E_N=|\Delta E_N/E_N^{\mathrm{baseline}}|\times100\%$ of periodically modulated beating dynamics versus thermal occupation $\bar n$. Cryogenic ($\bar n\leq 0.05$) and moderate ($\bar n\leq 0.2$) regimes remain within 5\% and 20\% of the zero-temperature baseline, while higher occupancies show larger deviations confined to early transients and late-time decay.}
    \label{fig:beatingThermal}
\end{figure}

Finally, we test the stability of periodically driven beating against thermal excitations. Figure~\ref{fig:beatingThermal} reports the relative deviation of $E_N(t)$ from the zero-temperature baseline at $\delta_0/\kappa=-2$. In the cryogenic regime ($\bar n\leq0.05$), deviations remain below $5\%$, increasing to at most $20\%$ for moderate occupations ($\bar n\leq0.2$). At higher temperatures, the deviations can reach $100\%$, but these excursions are confined to early transients and late-time decay, while the central beating structure persists throughout. This robustness follows from the pseudomode embedding [Eqs.~\ref{eqn:pseudoHam}--\ref{eqn:pseudoEOM}], which reduces to the $\hat O_0$ closure at $\bar n=0$ [Appendix~\ref{app:Pseudo}] and guarantees complete positivity across $\bar n>0$. Consequently, periodic modulation preserves entanglement structure under experimentally accessible cryogenic conditions~\cite{pleasance_generalized_2020,luo_quantum-classical_2023,shiokawa_non-markovian_2009}.

\section{Discussion and Conclusion}\label{discussion}

Our results identify three mechanisms that do not arise in Markovian Gaussian channels. (i) An Ornstein--Uhlenbeck reservoir supports an analytic freezing condition in which a tuned detuning $\delta_{AE}/\kappa$ balances the real and imaginary parts of $F_{1,2}(\infty)$, collapsing trajectories across memory strengths to within a few percent. (ii) Structured reservoirs generate entanglement revivals from separable orthogonal inputs, contrasting with earlier work in which revivals required initial entanglement~\cite{maniscalco_entanglement_2007,shiokawa_non-markovian_2009,paz_dynamical_2009,vasile_continuous-variable-entanglement_2009}. (iii) Periodic detuning produces integer-locked beating, explained by Jacobi--Anger harmonic selection~\cite{gonzalez-henao_generation_2015,liao_reservoir-engineered_2018}. Together, these effects place bath memory as an active entanglement resource.

The Markovian baseline (Figs.~\ref{fig:markovBaseline},\,\ref{fig:markovThermal}) recovers the established bright--dark mode structure: aligned squeezing loads a decoherence-free dark mode, while orthogonal inputs remain separable~\cite{braun_creation_2002,prauzner-bechcicki_two-mode_2004,lidar_decoherence-free_2003,chou_exact_2008}. Incorporating OU correlations preserves complete positivity through the pseudomode embedding (Sec.~\ref{sec:pseudo}), enabling closed-form steady-state coefficients that capture freezing (Figs.~\ref{fig:freezingZeroT},\,\ref{fig:freezingThermal}) and orthogonal-input birth–death–revival cycles (Fig.~\ref{fig:orthogonalRevivals}) consistent with phase accumulation and kernel feedback~\cite{paz_dynamical_2009,horhammer_environment-induced_2008}. Periodic modulation selects discrete harmonics (Figs.~\ref{fig:integerBeating}--\ref{fig:integerLocking}), yielding long-lived square-wave revivals unique to structured reservoirs.

Two limitations define the scope. The non-Markovian witness $\mathcal N$~\cite{breuer_measure_2009,rivas_quantum_2014,banchi_quantum_2015} is not maximized over inputs; thus $\mathcal N=0$ signifies only nondetection, not absence of memory. Thermal effects introduce no new mechanisms: occupation rescales symplectic spectra by $\sqrt{1+2\bar n}$, producing monotonic suppression consistent with Gaussian-channel studies~\cite{paz_dynamics_2008,dumitru_entanglement_2015,horhammer_environment-induced_2008,wolf_entangling_2011}. Deviations remain $\le 5\%$ for $\bar n\le0.05$ and $\le 20\%$ for $\bar n\le0.2$ (Figs.~\ref{fig:freezingThermal},\,\ref{fig:beatingThermal}), matching contemporary cavity, phononic, and optomechanical platforms~\cite{gonzalez-henao_generation_2015,bai_amplitude-modulation-based_2019,liao_reservoir-engineered_2018,woolley_two-mode_2014,park_single-mode_2024,ockeloen-korppi_stabilized_2018}. The pseudomode method~\cite{barnett_hazards_2001,pleasance_generalized_2020,luo_quantum-classical_2023,aolita_open-system_2015} ensures complete positivity at all temperatures and supports extensions to multimode and hybrid reservoirs.

Future directions include testing universality of freezing across higher-order kernels and multimode architectures~\cite{direkci_universality_2025,teklu_continuous-variable_2022,fiurasek_gaussian_2012}, incorporating non-Gaussian and controlled drives~\cite{mui_enhanced_2025,chen_optimal_2025}, and refining diagnostics to tighten the relationship between entanglement dynamics and memory via optimized Gaussian fidelities~\cite{bittel_optimal_2025,banchi_quantum_2015}. These results establish structured reservoirs as tunable, analytically tractable entanglement resources for continuous-variable platforms.

\appendix

\section{Gaussian Baseline and Symplectic Criteria}\label{app:BDSS}

\subsection{Bright-Dark Mode Decomposition}

The system couples to the reservoir through the collective operator of Sec.~\ref{sec:sysdef}, which acts symmetrically on both modes. We make this structure explicit by introducing bright and dark modes through a unitary rotation of the ladder operators \cite{chou_exact_2008,zhao_density_2002,ferraro_gaussian_2005} 

\begin{equation}\label{eqn:BDOps}
    \hat{d}_{\pm} = \frac{1}{\sqrt{2}}(\hat{a}\pm\hat{b}).
\end{equation} 

In this basis, the free Hamiltonian separates into number terms and cross terms proportional to the mode detuning

\begin{equation}\label{eqn:BDHam}
    \hat{H}_{BD} = \hbar(\omega_{a}+\frac{\delta_{AB}}{2})(\hat{d}^{\dagger}_{+}\hat{d}_{+} + \hat{d}^{\dagger}_{-}\hat{d}_{-}) - \hbar\frac{\delta_{AB}}{2}(\hat{d}^{\dagger}_{+}\hat{d}_{-} + \hat{d}^{\dagger}_{-}\hat{d}_{+}).
\end{equation}

The dark mode decouples dynamically only at resonance, when the cross terms vanish \cite{lidar_decoherence-free_2003,prauzner-bechcicki_two-mode_2004,horhammer_environment-induced_2008}. The dissipative channel also simplifies: the jump operator reduces to the bright mode

\begin{equation}\label{eqn:BDJump}
     \hat{L} = \sqrt{2\kappa}\hat{d}_{+},
\end{equation}

leaving the dark mode isolated from the bath \cite{braun_creation_2002,benatti_environment_2003}.

At resonance, the dark mode is free of both Hamiltonian and dissipative dynamics, forming a decoherence-free subspace \cite{lidar_decoherence-free_2003}. In this scenario, the bright mode decoheres into the vacuum, while the dark mode becomes a displaced squeezed thermal state. Utilizing the equations of motion outlined in Sec.~\ref{sec:theory} with $\delta_{AB} = 0$ yields the following dark mode steady state

\begin{align}
    \langle \hat d_{-} \rangle_{\infty} &= \frac{\alpha - \beta}{\sqrt{2}}, \\
    \langle \hat d_{-}^2 \rangle_{\infty} &= \frac{(\alpha - \beta)^2}{2} - \frac{1}{4}\sum_{o \in \{a,b\}}\mathrm{sgn}(s_o)\sinh[2|s_o|], \\
    \langle \hat d_{-}^\dagger \hat d_{-} \rangle_{\infty} &= \frac{|\alpha - \beta|^2}{2} + \frac{1}{2}\sum_{o \in \{a,b\}}\sinh[|s_o|]^2,
\end{align}

where $s_o \in \mathbb{R}$ and $\mathrm{sgn}(s_o)$ returns the sign of $s_o$. This structure explains the Markovian baseline in Fig.~\ref{fig:markovBaseline}: aligned (same signed) inputs introduce squeezing into the dark mode by updating both $\langle \hat d_{-}^2 \rangle_{\infty}$ and $\langle \hat d_{-}^\dagger \hat d_{-} \rangle_{\infty}$, while orthogonal (oppositely signed) inputs reduce the squeezing as the summation in $\langle \hat d_{-}^2 \rangle_{\infty}$ becomes a subtraction. The thermalization is still preserved fully in the dark mode, reducing the entanglement resource \cite{adesso_entanglement_2007,ferraro_gaussian_2005}.

\subsection{Simon Separability Criteria}

Entanglement in Gaussian states is certified by the Peres–Horodecki partial transpose condition, which has an explicit covariance-matrix form \cite{simon_peres-horodecki_2000,duan_inseparability_2000,adesso_entanglement_2007}. For the squeezed-vacuum inputs studied here, the pairing of a vacuum bright mode and a displaced squeezed thermal dark mode allows one to express the relevant steady state dynamics of the original modes as:

\begin{align}
    \langle \hat a \rangle_{\infty} &= -\langle \hat b \rangle_{\infty} =  \frac{\alpha - \beta}{2}, \\
    \langle \hat a^2 \rangle_{\infty} &= \langle \hat b^2 \rangle_{\infty} \nonumber \\
    &= -\langle \hat a \hat b \rangle_{\infty} \nonumber\\
    &=\frac{1}{4}(\alpha-\beta)^2 - \frac{1}{8}\sum_{o\in\{a,b\}}\mathrm{sgn}(s_o)\sinh[2|s_o|],\\
    \langle \hat a^\dagger \hat a \rangle_{\infty} &= \langle \hat b^\dagger \hat b \rangle_{\infty}\nonumber \\
    &= -\mathrm{Re}[\langle \hat a \hat b^\dagger \rangle_{\infty}] \nonumber \\
    &= \frac{1}{4}|\alpha-\beta|^2 + \frac{1}{4}\sum_{o \in \{a,b\}}\sinh^2[|s_o|], \\
    \mathrm{Im}&[\langle \hat a \hat b^\dagger \rangle_{\infty}] = 0,
\end{align}

where the same notation as the bright-dark mode is used. The solutions can be used to define the steady-state covariance matrix, as outlined in Eqs.~\eqref{eqn:quadOps}-\eqref{eqn:quadOrder}, to give 

\begin{equation}\label{eqn:steady_covariance}
    \sigma_{\infty} = \frac{1}{8}\begin{pmatrix}
        A_{-} && \mathbf{0}\\
        \mathbf{0} && A_{+}
    \end{pmatrix}
\end{equation}

where $\mathbf{0}$ is a $2 \times 2$ zero matrix, and $A_{\pm}$ is the remaining covariance structure 

\begin{equation}
    A_{\pm} = \begin{pmatrix}
        2 + e^{\pm2s_a} + e^{\pm2s_b} && 2 - e^{\pm2s_a} - e^{\pm2s_b}\\
        2 - e^{\pm2s_a} - e^{\pm2s_b} && 2 + e^{\pm2s_a} + e^{\pm2s_b}
    \end{pmatrix},
\end{equation}

which is only dependent on the individual mode squeezing ($s_o$) with coherent displacements ($\alpha, \beta$) removed by the covariance definition given in Eq.~\eqref{eqn:covMat}. Because $A_{\pm}$ depends only on $s_a$ and $s_b$, all symplectic invariants entering the Simon criterion and logarithmic negativity depend only on the input squeezing parameters. For \textbf{positively aligned squeezings}, inseparability requires the sum of variances to remain below the vacuum threshold,

\begin{equation}\label{eqn:SimonPos}
    e^{-2s_a}+e^{-2s_b}<2; \qquad (s_a>-\tfrac{\ln 2}{2},\,s_b>-\tfrac{\ln 2}{2}),
\end{equation}

while for \textbf{negatively aligned squeezings}, the complementary condition applies,

\begin{equation}\label{eqn:SimonNeg}
    e^{2s_a}+e^{2s_b}<2; \qquad (s_a<\tfrac{\ln 2}{2},\,s_b<\tfrac{\ln 2}{2}),
\end{equation}

These specialized forms of the Simon criterion delimit the entangled quadrants in squeezing space \cite{simon_peres-horodecki_2000,duan_inseparability_2000}.

Evaluating them gives explicit cutoffs. For instance, fixing $s_a=+1$ requires $s_b>-0.312$, consistent with the numerical threshold in Fig.~\ref{fig:markovBaseline}(a). The same conditions generate the red contours in Fig.~\ref{fig:markovBaseline}(b), which enclose the aligned quadrants where steady entanglement persists. Orthogonal and anti-aligned inputs fall outside these regions and decay to separable vacua under Markovian dynamics \cite{adesso_entanglement_2007,ferraro_gaussian_2005}.

\subsection{Symplectic Eigenvalue Formulas}

Entanglement in Gaussian systems is diagnosed by the spectrum of the partially transposed covariance matrix. Separability holds when the smallest symplectic eigenvalue satisfies $\tilde\nu_{-}\geq \tfrac{1}{2}$ \cite{simon_peres-horodecki_2000}. This condition underlies logarithmic negativity, a computable monotone defined directly from $\tilde\nu_{-}$ \cite{vidal_computable_2002,adesso_entanglement_2007}.

For the squeezed-coherent states of this study, the eigenvalues reduce to closed forms depending only on local squeezings as defined in the steady state covariance of Eq.~\eqref{eqn:steady_covariance} \cite{adesso_entanglement_2007}:

\begin{align}\label{eqn:sympleticEigens}
    \tilde\nu_1&=\frac{\sqrt{(e^{2s_a}+e^{2s_b})}}{2\sqrt2}\,e^{-(s_a+s_b)} \\
\tilde\nu_2&=\frac{\sqrt{(e^{2s_a}+e^{2s_b})}}{2\sqrt2}.
\end{align}

These formulas show that the thermal excitations rescale the spectrum uniformly as

\begin{equation}\label{eqn:themalEigens}
    \tilde{\nu}_{j}(\bar{n}) = \sqrt{1+2\bar{n}}\,\tilde{\nu}_{j},
\end{equation}

shifting the separability threshold while preserving inequality structure.

In practice, inseparability is set by $\tilde\nu_-=\min(\tilde\nu_1,\tilde\nu_2)$. These expressions give the analytic cutoffs shown in Fig.~\ref{fig:markovThermal}: aligned inputs support entanglement up to $\bar n\approx1.5$, while weakly orthogonal inputs lose correlations near $\bar n\approx0.2$. The agreement with numerical results in Sec.~\ref{subsec:markovBaseline} confirms that thermal suppression follows directly from the rescaled spectrum. These eigenvalue formulas thus provide the quantitative link between separability criteria and the observed finite-temperature decay of correlations \cite{adesso_entanglement_2007,weedbrook_gaussian_2012}.

\section{$O_0$ Formalism and Freezing Law}\label{app:O0Theory}

\subsection{$O_0$ Coefficient Dynamics}

The $\hat O_0$ closure gives a tractable description of finite-memory feedback from the Ornstein–Uhlenbeck kernel, extending the master equation of Sec.~\ref{sec:oZero}. Eqs.~\ref{eqn:oZeroAnsatz}–\ref{eqn:oZeroEOM} define the ansatz and averaged operator; here we connect the coefficient functions $f_i(t,s)$ to the convolution variables $F_i(t)$ through

\begin{equation}\label{eqn:convolutionCoeffs}
    F_{i}(t) = \int_{0}^tf_{i}(t,s)\alpha(t,s)ds, \qquad i\in\{1,2\}.
\end{equation}

This definition accumulates the reservoir kernel against the instantaneous closure coefficients.

The evolution of $\hat O_0(t,s)$ follows the deterministic closure equation \cite{yu_non-markovian_1999}

\begin{equation}\label{eqn:closure}
    \tfrac{\partial}{\partial t}\hat{O}(t,s) = -\tfrac{i}{\hbar}[\hat{H}_{S},\hat{O}_0(t,s)] - [\hat{L}^{\dagger}\hat{\bar{O}}_0(t),\hat{O}_0(t,s)],
\end{equation}

subject to the boundary condition $\hat O_0(t,t)=\hat L$. This fixes the initial coefficients as $f_i(t,t)=\sqrt{\kappa}$. Differentiating the convolution relation gives 

\begin{equation}\label{eqn:OURicatti}
    \dot{F}_{i}(t) = \tfrac{\gamma\sqrt{\kappa}}{2} -(\gamma+i\Omega)F_i(t) +\int_0^t\alpha(t-s)\tfrac{\partial}{\partial t}f_i(t,s)ds,
\end{equation}

where the single-exponential OU kernel allows recovery of the $F_i$ term.

Combining this identity with the closure dynamics reduces directly to the Riccati-type system of Eq.~\ref{eqn:consistency}, establishing the coefficient evolution used in the structured-bath analysis. This appendix supplies the derivation omitted in Methods.

The sequence from closure to convolution to Riccati form provides the analytic basis for the freezing overlap of Fig.~\ref{fig:freezingZeroT} and the revival structure of Fig.~\ref{fig:orthogonalRevivals} \cite{yu_non-markovian_1999,diosi_non-markovian_1998,barnett_hazards_2001}.

\subsection{Mode Resonant Steady-State}

At resonance ($\delta_{AB}=0$), the coefficient equations of motion [Eq.~\ref{eqn:consistency}] simplify and admit both a transient solution and a universal steady-state limit. The steady-state form $F_{1,2}(\infty)$ and $\chi(\gamma,\delta_{AE})$ are given in Methods Eqs.~\ref{eqn:modeResonanceSS}, \ref{eqn:SSFuncts}. The transient solution was used to plot analytic trajectories for resonant freezing, avoiding reliance on numerical integration.

The real part of $\chi$ sets the damping toward the steady state, while the imaginary part fixes the oscillation frequency. Choosing the branch with $\Re\chi>0$ ensures stability, so the transient decays monotonically to $F_{1,2}(\infty)$. This explains why resonant trajectories overlap regardless of reservoir correlation time.

The transient form makes the $\gamma$-dependence explicit and was used to generate the overlapping trajectories in Fig.~\ref{fig:freezingZeroT}(a). The steady-state solution accounts for the collapse of long-time curves in Fig.~\ref{fig:freezingZeroT}(b), where deviations remain bounded as correlation time varies. This analytic structure underlies the universal freezing behavior, consistent with broader non-Markovian analyses \cite{maniscalco_entanglement_2007,vasile_continuous-variable-entanglement_2009}, and agrees with numerical verification in App.~\ref{app:Numerics}.

\subsection{Freezing Law and Stability}

At exact resonance ($\delta_{AB}=0$), the convolution coefficients admit analytic inspection. Methods already list $F_{1,2}(\infty)$  Eq.~\ref{eqn:modeResonanceSS}, $\chi(\gamma,\delta_{AE})$  Eq.~\ref{eqn:SSFuncts}, and the freezing inequalities  Eqs.~\ref{eqn:environmentFreezing}, \ref{eqn:freezingMemoryBound}. Here we sketch how these results follow from the transient solution and show why they are stable. These derivations underlie the trajectory overlap in Fig.~\ref{fig:freezingZeroT} and the thermal robustness of Fig.~\ref{fig:freezingThermal}.

The resonant Riccati system yields a closed-form transient for the feedback coefficient

\begin{equation}\label{eqn:coeffResonantSolution}
    F(t)=\tfrac{1}{4\sqrt{\kappa}}\Big[(\gamma+i\delta_{AE})-\chi\,\tanh(\tfrac{1}{2}\chi t+\phi)\Big],
\end{equation}

with phase $\phi=\tan^{-1}[(\gamma+i\delta_{AE})/\chi]$. In the long-time limit, it reduces to $F_{1,2}(\infty)$. The analytic form makes the $\gamma$-dependence explicit and was used to generate the overlapping trajectories in Fig.~\ref{fig:freezingZeroT}(a).

Freezing is enforced by requiring correlations to decay to $1/n$ of their initial value at time $t_n$, giving

\begin{equation}\label{eqn:quantumDecay}
    e^{-4\sqrt{\kappa}F_{1,2}(\infty)t_n} = \frac{1}{n}.
\end{equation}

Substituting $F_{1,2}(\infty)$ yields the critical detuning $\delta_{AE}^*$ that cancels memory dependence. A second requirement demands faster convergence of the coefficients, defined by $(t_s,n_s)$. This produces a lower bound on $\gamma$ through $\chi$

\begin{equation}\label{eqn:coeffDecay}
    e^{-\Re(\chi)t_s} = \frac{1}{n_s}.
\end{equation}

Together, these conditions form the freezing law, explaining the $\gamma$-independent overlap of Fig.~\ref{fig:freezingZeroT}(b) and its persistence at finite $T$ in Fig.~\ref{fig:freezingThermal} \cite{breuer_measure_2009,rivas_quantum_2014}.

Stability follows by linearizing the Riccati equations around $F_{1,2}(\infty)$. The Jacobian eigenvalues have real part $-\Re\chi$, so choosing $\Re\chi>0$ ensures asymptotic stability. This decay rate coincides with the convergence bound above, showing consistency between analytic inequalities and dynamical stability \cite{barnett_hazards_2001}. Numerical confirmation is given in App.~\ref{app:Numerics}.

The combined decay criterion, convergence bound, and stability condition guarantee that any $(\gamma,\delta_{AE}^*)$ pair satisfying these relations yields indistinguishable entanglement trajectories. This establishes the universality of the freezing law across correlation times and finite temperatures.

\subsection{Detuning-Based Phase Accumulation}

A direct demonstration of phase accumulation follows from the detuning term in the coefficient equations Eq.~\ref{eqn:consistency}. Substituting $F_{2}(t)\to e^{i\delta_{AB}t}\tilde F_{2}(t)$ isolates the oscillatory part and yields an evolution equation with a time-dependent source and a global phase.

\begin{align}
    \tfrac{d}{dt}\tilde{F}_2(t) = \frac{\gamma\sqrt{\kappa}}{2}e^{-i\delta_{AB}t} + \\(\sqrt{\kappa}(F_1 +& e^{i\delta_{AB}t}\tilde{F}_2)-(\gamma+i\delta_{AE}))\tilde{F}_2.\nonumber
\end{align}
    
In the $\omega_a$ rotating frame, the same structure appears at the operator level: mode $b$ acquires a phase factor $e^{-i\delta_{AB}t}$ while mode $a$ remains fixed. These oscillations interfere with the Ornstein–Uhlenbeck kernel, alternately enhancing and suppressing the feedback. The result is a cycle of entanglement growth, decay, and revival. This explains Fig.~\ref{fig:orthogonalRevivals}, where revivals occur only in structured baths and vanish in the Markovian baseline.

Earlier studies reported revivals from initially entangled states \cite{maniscalco_entanglement_2007,shiokawa_non-markovian_2009,paz_dynamical_2009,vasile_continuous-variable-entanglement_2009}. Here they emerge from separable squeezed inputs, extending the Markovian baselines of Prauzner-Bechcicki \cite{prauzner-bechcicki_two-mode_2004} and Hörhammer \cite{horhammer_environment-induced_2008}. Detuning and reservoir memory together generate long-lived correlations from uncorrelated modes, and the same phase mechanism underlies the beating effects analyzed in Sec.~\ref{subsec:beating}.

\section{Periodic Modulation Analysis}\label{app:PMA}

In the sinusoidally driven case the detuning is $\delta_{AB}(t)=\delta_{0}(\sin t+1)$. Phase accumulation then extends the constant-detuning analysis by introducing a time-dependent modulation of the $b$-mode sector. The coefficient $F_{2}(t)$ acquires an additional oscillatory factor from the integrated detuning

\begin{equation}\label{eqn:phaseAccum} e^{\pm i \int_{0}^{t}\delta_{AB}(s)ds} = e^{\pm i\frac{\delta_{0}}{\omega}}e^{\pm i \delta_{0}t}e^{\mp i\frac{\delta_{0}}{\omega}\cos(\omega t)}, \end{equation}

where we update to $\delta_{AB}(t)=\delta_{0}(\sin \omega t+1)$ and fix $\omega/\kappa = 1$ in the main text. Unlike the constant case, where the averaged phase cancels at large detuning, this modulation produces structured oscillations that seed the beating envelopes of Fig.~\ref{fig:integerBeating}. The effect has no analog in the Markovian baseline \cite{prauzner-bechcicki_two-mode_2004,horhammer_environment-induced_2008,yu_non-markovian_1999,barnett_hazards_2001}.

The oscillatory factor can be decomposed using the Jacobi–Anger identity:  \begin{equation}\label{eqn:jacobi}
    e^{\pm i \delta_{0}t}e^{\mp i\frac{\delta_0}{\omega}\cos(\omega t)} = \sum_{k \in \mathbb{Z}}i^{k}J_{k}(\mp \frac{\delta_{0}}{\omega})e^{i(k\omega\pm\delta_0)t},
\end{equation}

giving a sum of Bessel-weighted harmonics. Long-time averaging selects commensurate terms with $\delta_{0}/\omega=k\in\mathbb Z$, yielding the integer-locking condition. When $\omega/\kappa = 1$, this reduces to $\delta_0/\kappa \in \mathbb{Z}$. This explains the sharp peaks in the time-averaged entanglement of Fig.~\ref{fig:integerLocking}(a) and is consistent with prior studies of modulation-induced correlations in structured reservoirs \cite{gonzalez-henao_generation_2015,bai_amplitude-modulation-based_2019,liao_reservoir-engineered_2018,woolley_two-mode_2014,hsiang_entanglement_2022,yang_quantum_2025}.

The integer-locking rule provides the analytic basis for the square-wave envelopes in Fig.~\ref{fig:integerBeating}, the entanglement peaks in Fig.~\ref{fig:integerLocking}(a), and their persistence under cryogenic noise in Fig.~\ref{fig:beatingThermal}. These features vanish in the Markovian limit, confirming that modulation-induced phase selection requires reservoir memory. Finite-temperature robustness follows from the pseudomode construction of Sec.~\ref{sec:pseudo}, which guarantees complete positivity of the driven dynamics. Together, these results connect number-operator detuning modulation to integer-locked beating in structured environments, extending reservoir-control strategies and placing the effect within experimental reach of cavity and phononic platforms \cite{krauter_entanglement_2011,ockeloen-korppi_stabilized_2018,park_single-mode_2024,semmler_single-mode_2016,aolita_open-system_2015,breuer_measure_2009,rivas_quantum_2014}.

\section{Pseudomode Embedding}\label{app:Pseudo}

Starting from the pseudomode embedding defined in Eqs.~\ref{eqn:pseudoHam} and \ref{eqn:pseudoJumps}, the Heisenberg equation for $\hat c$ can be derived using Eq.~\ref{eqn:pseudoEOM}, 

\begin{equation}\label{eqn:pseudoCmode}
    \tfrac{d}{dt} \hat{c}  = -(\gamma+i\Omega)\hat{c} -g\sqrt{\kappa}( \hat{a} + \hat{b}),
\end{equation}

isolating homogeneous decay with rate $\gamma$ and oscillation frequency $\Omega$. Its formal solution is obtained with an integrating factor,

\begin{align}\label{eqn:pseudoCgeneral}
    \hat{c}(t) = e^{-(\gamma+i\Omega)(t-s)}\hat{c}(s)& \\ - i\,g\int_{s}^{t}&e^{-(\gamma+i\Omega)(t-\tau)}(\hat{a}(\tau)+\hat{b}(\tau))d\tau\nonumber,
\end{align}

yielding an exponential damping term multiplied by the initial condition plus a convolution with the collective operator. Eliminating $\hat c$ from the $(a,b)$ equations of motion produces a memory integral with kernel $g^{2}e^{-(\gamma+i\Omega)(t-\tau)}$. Choosing $g=\sqrt{\gamma/2}$ recovers the Ornstein--Uhlenbeck form given in Eq.~\ref{eqn:kernel}, confirming that the pseudomode embedding is algebraically equivalent to the convolution kernel. This correspondence follows the standard pseudomode construction for structured reservoirs \cite{garraway_nonperturbative_1997,pleasance_generalized_2020,luo_quantum-classical_2023}.

The Lindblad operators in Eq.~\ref{eqn:pseudoJumps} drive the pseudomode into a stationary thermal state with $\langle \hat c^\dagger \hat c\rangle=\bar n$ and $\langle \hat c \rangle=0$, which must be chosen as the initial conditions, $\langle \hat{c} \rangle_{t=0} = 0$ and  $\langle \hat{c}^{\dagger}\hat{c} \rangle_{t=0} = \bar{n}$, to avoid spurious transients. 

The corresponding two-time correlators retain the Ornstein--Uhlenbeck envelope but acquire thermal weights $(\bar n+1)$ and $\bar n$, so the functional form of the kernel remains unchanged while the balance between emission and absorption is adjusted. 

\begin{equation}
    \langle \hat{c}^{\dagger}(t)\hat{c}(s)\rangle = \bar{n}\alpha^{*}(t,s), \qquad
    \langle \hat{c}(t)\hat{c}^{\dagger}(s)\rangle = (\bar{n}+1)\alpha(t,s)
\end{equation}

In the zero-temperature limit $\bar n=0$, the absorption channel vanishes so that $L_{+}=0$, and the Heisenberg solution for $\hat c$ contains only emission noise. Eliminating $c$ under these conditions reproduces the Ornstein--Uhlenbeck correlator as, 

\begin{equation}
    \langle \hat{c}^{\dagger}(t)\hat{c}(s)\rangle = 0 \qquad
    \langle \hat{c}(t)\hat{c}^{\dagger}(s)\rangle = \alpha(t,s)
\end{equation}

The resulting nonlocal term in the $(a,b)$ dynamics coincides with the $\hat{\bar O}_0(t)$ closure introduced in Sec.~\ref{sec:oZero}, with the convolution coefficients $F_{1,2}(t)$ entering identically.  
Finally, the pseudomode embedding at $\bar n=0$ reduces exactly to the QSD closure, confirming that the finite-temperature and zero-temperature frameworks are two limits of a single consistent construction \cite{barnett_hazards_2001,garraway_nonperturbative_1997}.

Because the full tripartite system $(a,b,c)$ evolves under the GKLS master equation of Eq.~\ref{eqn:pseudoEOM}, tracing out the pseudomode yields a completely positive reduced map for $(a,b)$ at all times, ensuring physical consistency of the finite-temperature dynamics reported in Figs.~\ref{fig:freezingThermal} and \ref{fig:beatingThermal} \cite{horhammer_environment-induced_2008,shiokawa_non-markovian_2009,pleasance_generalized_2020}.

\section{Numerical Verification}\label{app:Numerics}

Across all parameter scans, halving the adaptive step size and tightening solver tolerances shifted results by less than the thresholds defined in Sec.~\ref{sec:numerical}, with $\max_t|\Delta E_N|<10^{-6}$, $\max_t|\Delta D_B|<10^{-6}$, and time-integrated changes $<10^{-5}$. These tests were applied in the freezing regime (Figs.~\ref{fig:freezingZeroT}--\ref{fig:freezingThermal}), orthogonal revivals (Fig.~\ref{fig:orthogonalRevivals}), and periodic modulation (Figs.~\ref{fig:integerBeating}--\ref{fig:beatingThermal}). No qualitative differences were observed among solver variants, and higher working precision was unnecessary, confirming that all reported trajectories are numerically stable and reproducible.

All reported trajectories satisfied the Gaussian physicality conditions stated in Sec.~\ref{sec:numerical}, including positivity of $\sigma+i\mathbf{J}/2$, minimum symplectic eigenvalue $\tilde\nu_{\min}\geq 1/2-10^{-4}$, and eigenvalue cutoffs. Parameter points that violated these bounds were excluded, ensuring that the logarithmic negativity and Bures metrics were never evaluated on nonphysical states.

Certain parameter points were excluded from production scans. At double resonance $\delta_{AB}=0=\delta_{AE}$ and $\gamma/\kappa=4$ under the $O_{0}$ closure, the analytic solution crosses from trigonometric to hyperbolic form, yielding the rational asymptote
\begin{equation}
    F(t)=\sqrt{\kappa}\frac{2\kappa t}{1+2\kappa t}.
\end{equation}
Numerical integration is unstable when $\gamma/\kappa < 4$ and $\delta_{AE} = 0$ simultaneously; therefore, production runs used $\delta_{AE}=0.1\kappa$ to represent \textit{near resonance} instead. In addition, the near-resonant Bures witness sector (Sec.~\ref{sec:numerical}) was excluded, while the Markov limit at $\gamma/\kappa=10$ remained stable.



\bibliographystyle{apsrev4-2}
\bibliography{RDGC}

\end{document}